\documentclass[12pt]{iopart}

\usepackage{graphicx}
\begin{document}

\title[Can transport peak explain the low-mass enhancement of dileptons at RHIC?]{Can transport peak explain the low-mass enhancement of dileptons at RHIC?}

\author{Y Akamatsu$^{1,2}$, H Hamagaki$^3$, T Hatsuda$^{4,2,5}$ and T Hirano$^{6,2}$}

\address{$^1$Kobayashi-Maskawa Institute for the Origin of Particles and the Universe (KMI), Nagoya University, Nagoya 464-8602, Japan}
\address{$^2$Department of Physics, The University of Tokyo, Tokyo 113-0033, Japan}
\address{$^3$Center for Nuclear Study, The University of Tokyo, Tokyo 113-0033, Japan}
\address{$^4$Theoretical Research Division, Nishina Center, RIKEN, Wako 351-0198, Japan}
\address{$^5$Institute for the Physics and Mathematics of the Universe (IPMU), The University of Tokyo, Kashiwa 277-8568, Japan}
\address{$^6$Department of Engineering and Applied Science, Sophia University, Tokyo 102-8554, Japan}

\ead{akamatsu@kmi.nagoya-u.ac.jp}

\begin{abstract}
We propose a novel relation between the low-mass enhancement of dielectrons observed at PHENIX and transport coefficients of QGP such as the charge diffusion constant $D$ and the relaxation time $\tau_{\rm J}$. 
We parameterize the transport peak in the spectral function using the second-order relativistic dissipative hydrodynamics by Israel and Stewart.
Combining the spectral function and the full (3+1)-dimensional hydrodynamical evolution with the lattice EoS, theoretical dielectron spectra and the experimental data are compared.
Detailed analysis suggests that the low-mass dilepton enhancement originates mainly from the high-temperature QGP phase where there is a large electric charge fluctuation as obtained from lattice QCD simulations.
\end{abstract}


\section{Introduction}
The property of quark-gluon plasma (QGP) has been studied in relativistic heavy ion collision experiments at the Relativistic Heavy Ion Collider (RHIC) and at the Large Hadron Collider (LHC).
The discovery of large elliptic flow $v_2$ at RHIC, as large as hydrodynamical prediction with nearly perfect fluid, has revealed a new property of the QGP: strongly interacting nature of QGP \cite{Hirano:2008hy}.

Electromagnetic radiation, i.e. photons and dileptons, from matter under dynamical evolution carries clean information on the created hot matter.
In experiment, the other sources for the electromagnetic radiation, such as Drell-Yan process and hadronic decays, can be estimated so that the radiation from the dynamical medium can be isolated.
Recently, PHENIX Collaboration at RHIC found a low-mass enhancement of the dielectron spectra \cite{Adare:2009qk}, which is attributed to the radiation from dynamical medium.
We study the connection between the low-mass enhancement and transport peak in the spectral function by utilizing the medium profile calculated by the full (3+1)-dimensional hydrodynamic model with the lattice equation of state (EoS) \cite{Borsanyi:2010cj}.

\section{Transport Peak in Spectral Function}
Radiation spectrum of a lepton pair with momentum $p_1$ and $p_2$ from a unit spacetime volume of thermal medium with temperature $T$ is given by
\begin{eqnarray}
\frac{E_1E_2dN_{l^+l^-}}{d^3p_1d^3p_2d^4x}
&=&\frac{\alpha^2}{2\pi^4q^4}\frac{p_1^{\mu}p_2^{\nu}+p_2^{\mu}p_1^{\nu}-\frac{q^2}{2}g^{\mu\nu}}{\exp (q^0/T)-1}
{\rm Im}G_{\rm R}^{\mu\nu}(q;T),\\
G_{\rm R}^{\mu\nu}(q;T)&\equiv& \int d^4x e^{iqx}i\theta(x^0)\langle[J^{\mu}(x),J^{\nu}(0)]\rangle_T,
\end{eqnarray}
where 
$J^{\mu}(x)\equiv
\frac{2}{3}\bar {\rm u}\gamma^{\mu}{\rm u}
-\frac{1}{3}\bar {\rm d}\gamma^{\mu}{\rm d}
-\frac{1}{3}\bar {\rm s}\gamma^{\mu}{\rm s}$
is the electric current of quarks, 
$G_{\rm R}^{\mu\nu}(q;T)$ is the retarded Green function, and ${\rm Im}G_{\rm R}^{\mu\nu}(q;T)$ is the spectral function.
Previous phenomenological studies on dilepton production from heavy ion collisions at SPS and at RHIC suggest that novel production process is taking place at RHIC \cite{Rapp:2010sj}.
In these studies, the spectral function in hadronic phase has been calculated on the basis of hadronic many-body theory and/or chiral symmetry principles and that in QGP phase on the basis of perturbation theory.
Non-perturbative process such as transport process in QGP phase and crossover region has not been considered seriously and may be one of the candidates for the novel production process and we study it in detail.

Starting from Israel-Stewart theory \cite{Israel:1976tn,Israel:1979wp} for causal dissipative hydrodynamics in Landau frame with external electromagnetic field $\delta A^{\mu}(x)$ and performing linear analysis, we obtain linear equations for the external field and induced electric current $\delta J^{\mu}(x)=J^{\mu}(x)-J^{\mu}$ (for QGP with vanishing baryon chemical potential $J^{\mu}=(0,\vec 0)$):
\begin{eqnarray}
\delta J^{\mu}(x)=(\delta \rho (x),\delta \vec j(x)), \ 
\partial_{\mu}\delta J^{\mu}=0,\
\delta \vec j(x)=\sigma\delta\vec E
-D\vec\nabla\delta \rho
-\tau_{\rm J}\frac{\partial \delta\vec j}{\partial t},\\
D\equiv \frac{\sigma}{\chi},\ 
\tau_{\rm J} \equiv \beta_1\sigma, \ 
\chi \equiv \frac{\partial \rho}{\partial \mu},
\end{eqnarray}
where we introduce three transport coefficients $D$~(diffusion constant), $\sigma$~(conductivity), and $\tau_{\rm J}$~(relaxation time) and $\delta\vec E=-\vec \nabla \delta A^0-\partial_t \delta \vec A$ is the external electric field.
Applying the linear response relation $\langle\delta J^{\mu}\rangle=-G_{\rm R}^{\mu\nu}(q;T)\delta A_{\nu}(q)$, we obtain transport peaks in the two independent components of the spectral function:
\begin{eqnarray}
{\rm Im}G^{\rm (L)}_{\rm R}(q;T)
\equiv-\frac{q^2}{k^2}{\rm Im}G^{00}_{\rm R}(q;T)
=-\frac{\chi D \omega q^2}{\omega^2+(\tau_{\rm J}\omega^2-Dk^2)^2} \label{eq:SPF-T},\\
{\rm Im}G^{\rm (T)}_{\rm R}(q;T)
\equiv\frac{1}{2}\left({\rm Im}G^{\mu}_{\rm R,\mu}(q;T)
-{\rm Im}G^{\rm (L)}_{\rm R}(q;T)\right)
=-\frac{\chi D\omega}{\tau_{\rm J}^2\omega^2+1} \label{eq:SPF-L}.
\end{eqnarray}
Since dilepton rate is proportional to ${\rm Im}G^{\mu}_{\rm R,\mu}(q;T)/q^2$, the dilepton rate from transport peak at $\omega\sim |\vec k|\sim 0$ becomes divergent.
Setting $\vec k=\vec 0$, the transport peak is
${\rm Im}G^{\rm (L)}_{\rm R}={\rm Im}G^{\rm (T)}_{\rm R}=-\chi D\omega/(\tau_{\rm J}^2\omega^2+1)$.
The transport peak at fixed $\omega$ becomes stronger for larger $\chi$ and $D$ and for smaller $\tau_{\rm J}$.
The reason is as follows:
When $D$ is large and $\tau_{\rm J}$ is small, strong electric current is swiftly induced by the electric charge fluctuation, which is proportional to the susceptibility $\chi$.
This strong induced current emits a large number of photons and dileptons.
We parameterize the transport coefficients $D\propto 1/T$, $\tau_{\rm J}\propto 1/T$ for dimensional reason and the susceptibility  $\chi(T)=0.28T^2\left[1+\tanh\left(\frac{T[{\rm GeV}]-0.155}{0.023}\right)\right]$ by fitting a lattice result \cite{Allton:2005gk}.
We refer to the theoretical values of transport coefficients calculated by pQCD \cite{Arnold:2000dr, Arnold:2003zc, Hong:2010at} $(D^{\rm pQCD}T,\tau^{\rm pQCD}_{\rm J}T)=(4,15)$ and those by AdS/CFT \cite{Natsuume:2007ty} $(D^{\rm  AdS/CFT}T,\tau^{\rm AdS/CFT}_{\rm J}T)=(1/2\pi,\ln 2/2\pi)$.

\section{Dielectron Spectrum in Heavy Ion Collisions}
\begin{figure}
\begin{center}
\includegraphics[width=5cm, angle=-90, clip]{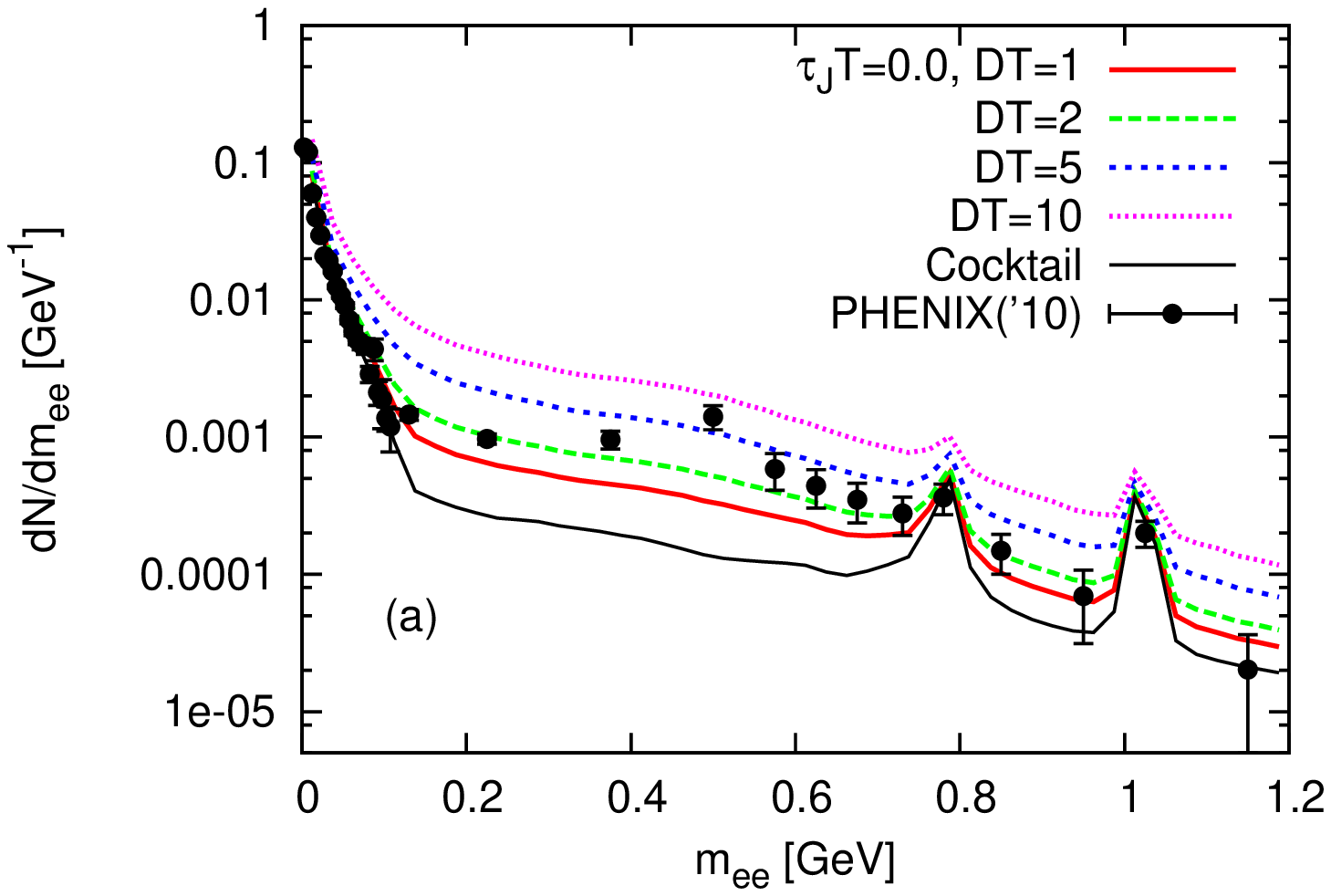}
\includegraphics[width=5cm, angle=-90, clip]{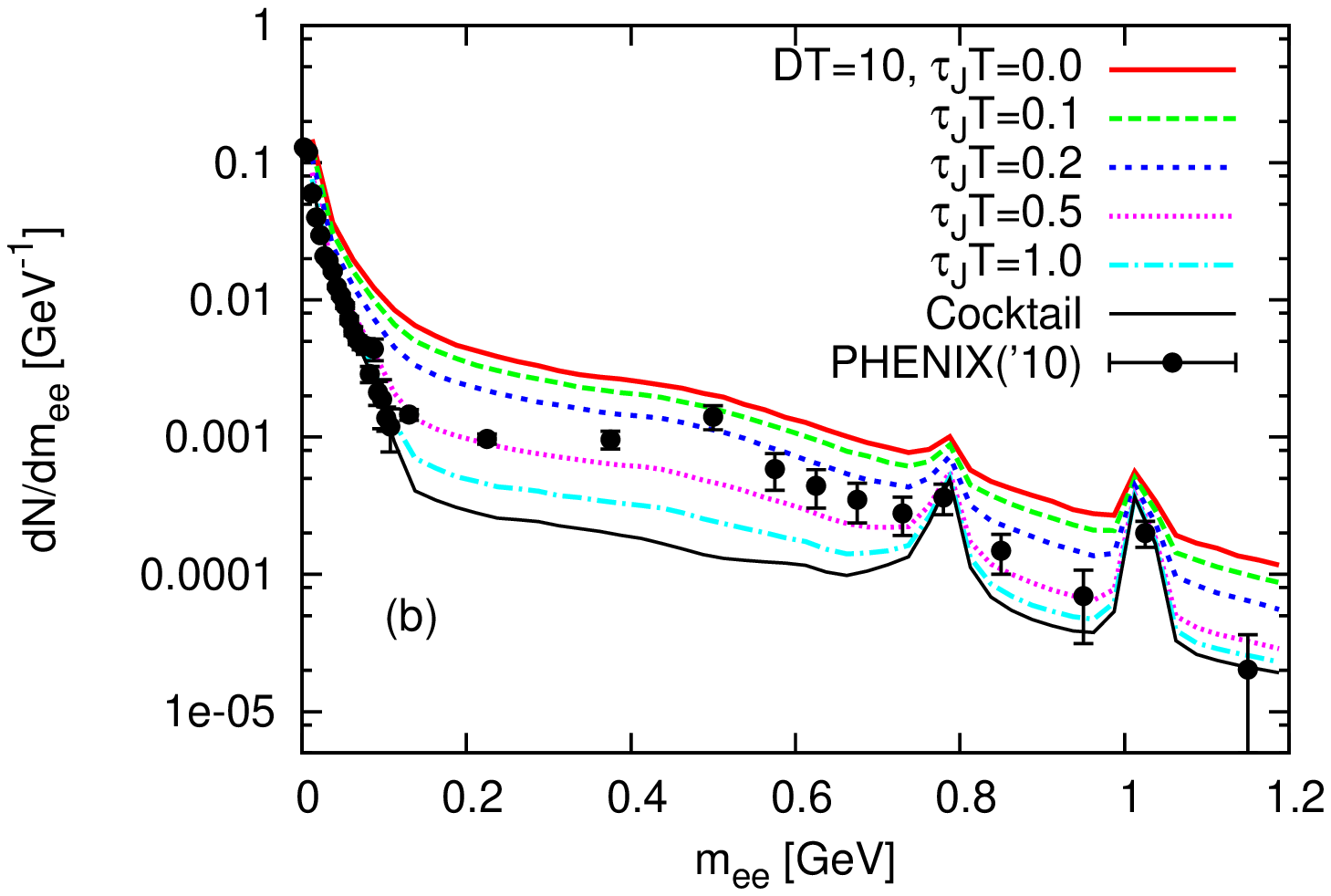}
\includegraphics[width=5cm, angle=-90, clip]{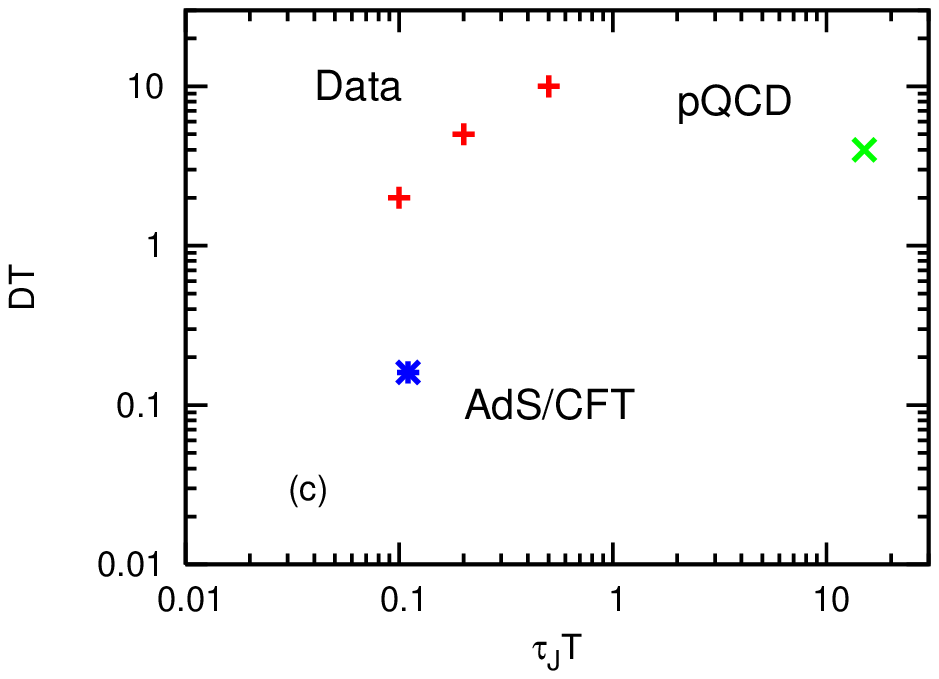}
\includegraphics[width=5cm, angle=-90, clip]{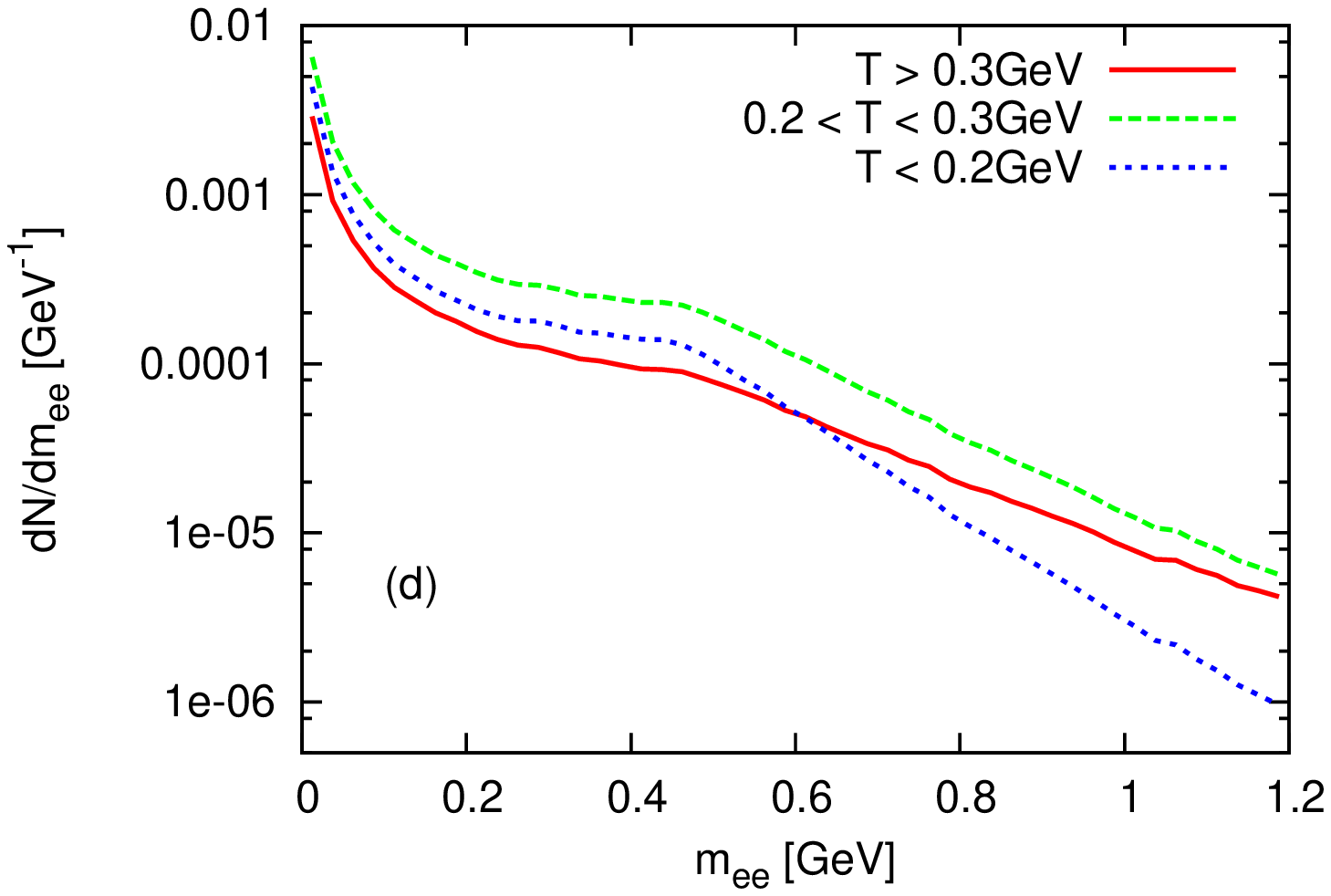}
\caption{
\footnotesize
(Color online)
The dielectron spectra from transport peak with (a) $\tau_{\rm J}=0$ and (b) $D=10/T$ fixed are compared with the experimental data \cite{Adare:2009qk}.
We take into account the contributions from the hadronic decays after freezeout (denoted as ``Cocktail'' in these figures).
In (c), the parameter sets ($D, \tau_{\rm J}$) with which the data can be fitted well are summarized.
In (d), we plot the spectra of dielectrons radiated from matter with different temperature ranges.
}
\label{fig}
\end{center}
\end{figure}
We combine the transport spectral function given in (\ref{eq:SPF-T}) and (\ref{eq:SPF-L}) with medium profile calculated by full (3+1)-dimensional hydrodynamic model with the lattice EoS \cite{Borsanyi:2010cj}.
For details of the hydrodynamic model adopted, see \cite{preparation} and references therein.
Shown in Fig.~\ref{fig}(a) are dielectron spectra for various $D$ and for $\tau_{\rm J}=0$.
Since spectral function with $\tau_{\rm J}=0$ gives the largest dielectron yield for each $D$, the data cannot be explained by $D<2/T$.
In Fig.~\ref{fig}(b), we show the dielectron spectra with various $\tau_{\rm J}$ and for $D=10/T$.
The dielectron spectrum with $(DT,\tau_{\rm J}T)=(10,0.5)$ shows good agreement with the data.
For each $D(\geq 2/T)$, we can find $\tau_{\rm J}$ with which the data can be fitted well.
We summarize in Fig.~\ref{fig}(c) some of such parameter sets $(DT,\tau_{\rm J}T)$ and theoretical estimates by pQCD and AdS/CFT.

The transport spectral function is proportional to $\chi$, which rapidly grows at high temperature.
The dilepton production is proportional not only to the spectral function but also to the spacetime volume, which is large at low temperature.
Therefore it is interesting to see the interplay between these competing effects, namely fluctuation and volume.
Shown in Fig.~\ref{fig}(d) is the spectra of dielectrons emitted from the matter with different temperature regions.
Clearly dielectrons are emitted dominantly from the matter with $0.2\leq T\leq 0.3$ GeV, suggesting that the fluctuation becomes important at high temperature.

\section{Conclusion}
We have calculated the dielectron yield by combining the transport peak and the medium evolution calculated by the full (3+1)-dimensional hydrodynamic model with the lattice EoS and attempted to extract the transport coefficients from the data.
We summarize our results in Fig.~\ref{fig}(c).
The obtained parameter sets are far from both weak-coupling pQCD calculation and strong-coupling calculation on the basis of the AdS/CFT correspondence.
We also showed that in order to explain the data, the diffusion coefficient must be $D\geq 2/T$.
In order to investigate the transport peak further, it is important to study the transverse momentum dependence of the dielectron spectra, which will be reported in \cite{preparation}.

\ack
Akamatsu~Y is partially supported 
by JSPS fellowships for Young Scientists.
Hatsuda~T is partially supported 
by No.~2004, Grant-in-Aid for Scientific Research on Innovative Areas.
Hirano~T is partially supported 
by Sumitomo Foundation No.~080734,
by Grant-in-Aid for Scientific Research No.~22740151.

\end{document}